\documentclass{acm_proc_article-sp}
\usepackage{amssymb}
\usepackage{graphicx}
\usepackage{url}
\usepackage{amsmath}
\usepackage{xcolor}

\title{Responding to Retrieval: A Proposal to Use Retrieval Information for Better Presentation of Website Content}

\begin{document}

\numberofauthors{3} 
%
\author{
%
%
\alignauthor
\small
C Ravindranath Chowdary\\
       \affaddr{\small IIT (BHU)}\\
       \affaddr{\small Varanasi, India 221 005}\\
       \email{\small rchowdary.cse@iitbhu.ac.in}
\alignauthor
\small
Anil Kumar Singh\\
       \affaddr{\small IIT (BHU)}\\
       \affaddr{\small Varanasi, India 221 005}\\
       \email{\small nlprnd@gmail.com}
\alignauthor
\small
Anil Nelakanti\\
       \affaddr{\small IIT (BHU)}\\
       \affaddr{\small Varanasi, India 221 005}\\
       \email{\small anil.nelakanti@gmail.com}
}

\maketitle


\begin{abstract}
Retrieval and content management are assumed to be mutually exclusive. In this paper we suggest that they need not be so. In the usual information retrieval scenario, some information about queries leading to a website (due to `hits' or `visits') is available to the server administrator of the concerned website. This information can used to better present the content on the website. Further, we suggest that some more information can be shared by the retrieval system with the content provider. This will enable the content provider (any website) to have a more dynamic presentation of the content that is in tune with the query trends, without violating the privacy of the querying user. The result will be a better synchronization between retrieval systems and content providers, with the purpose of improving the user's web search experience. This will also give the content provider a say in this process, given that the content provider is the one who knows much more about the content than the retrieval system. It also means that the content presentation may change in response to a query. In the end, the user will be able to find the relevant content more easily and quickly. 
\end{abstract}

\terms{Content Management Systems, Personalized Content, Information sharing}

\section{Introduction}
Information retrieval (IR) systems have become integral to daily activities of millions and will retain their prominence in years to come. One of the reasons for such importance of a good IR system is the amount of data that is available on the web and the pace at which it is increasing. The number of websites reportedly increased from one in 1991 to more than one billion in September 2014 \footnote{http://www.internetlivestats.com/total-number-of-websites/ on 27/10/2014}. Simultaneously, there was an increasing number of users availing the hosted services. This increase in web usage is more than an issue of load that was met by computationally powerful servers. The bigger challenge was to organize and make available the huge amount of information in a readily consumable manner. This required the third entity of retrieval systems. What essentially was a two-way transaction between the host and the client has become three-way with an IR system in the middle.

Clients are served by hosts, a relation facilitated by IR systems. However, current day IR systems are more than just organizer of web links. They model user choices and preferences to serve them better. We argue that the three-entity unit of the client, IR system and the host is greater than the sum of its parts. The relation between these three entities is ignored by the current web-service architecture. We present here a proposal which will exploit this relationship to better deliver some aspects for web service usage.

Web designers write content on the pages based on the information provided by the owner of the site. Content in a website is primarily organized based on the categorization of the information and arranged appropriately by the designer. In this entire process of the current design paradigm, the {\it query} has no role to play during the design or presentation phases of a website. But, when a search query is given to an IR system, it retrieves links of pages that are prepared without taking query into consideration {\it on the host side}. This is because retrieval and content management are considered mutually exclusive, that is, the content management system does not know about the retrieval system and the retrieval system does not know how the content provider may respond to the query. Due to this shortcoming, both the content provider and IR system are under performing. In this paper, we try to address this issue by proposing an architecture that enables the server hosting the website to present content that is based on the query posed by the user. 


\begin{figure*}[htb]
\centering
\epsfig{file=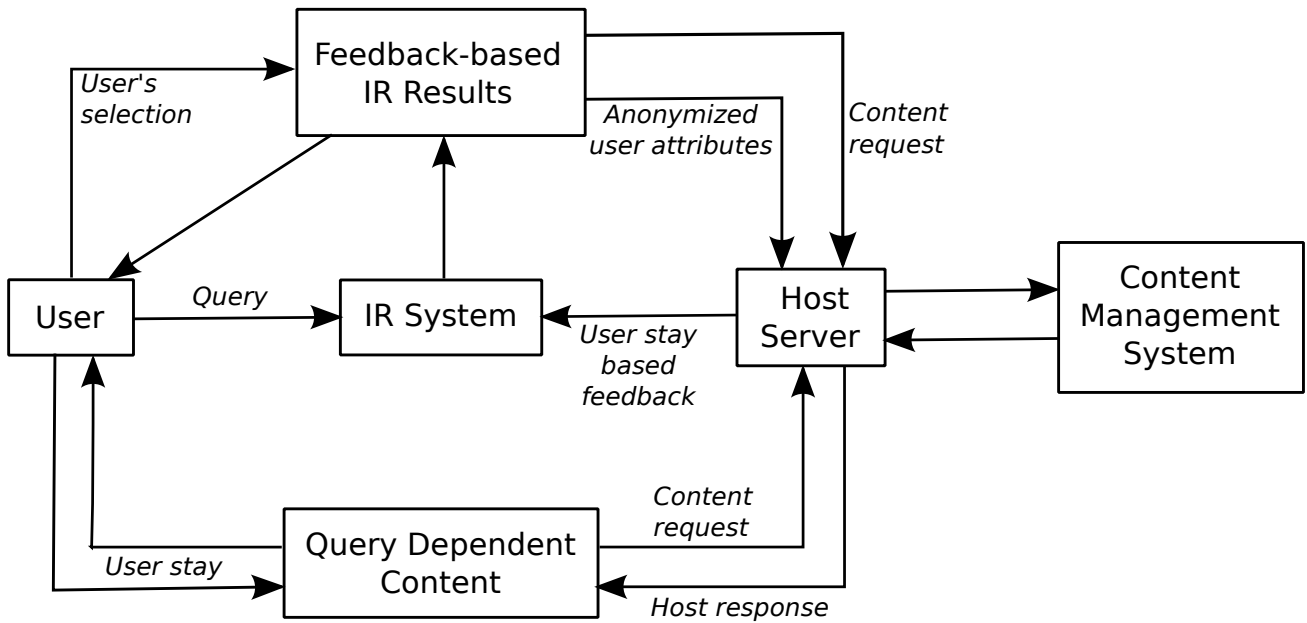, width=6.8in}
\caption{The proposed architecture for more responsive IR and CMS systems}
\end{figure*}

\section{The Proposed Architecture}

The outline of the architecture we propose is presented in Figure 1. The scenario is that the user starts a retrieval system and gives a query. The retrieval system presents the search results to the user. Out of them, the users selects one and is taken to the destination website. When the user is taken to that website, the retrieval system also shares some information about the user and the query (subject to privacy requirements: see Section 5) with the server hosting the website. The host server uses this information to present the content such that the user might have a better search experience (see Section 3). This presentation might, for example, make it easier for the user to find certain things. The host server will then provide feedback to the retrieval system (again subject to privacy requirements) based on the user's stay in the website and the user's activity during the stay. Since the information shared with the host server is anonymized, so will be the feedback given to the retrieval system. The retrieval system will now use this feedback to give better results in the future (see Section 4). The overall result will be better synchronization between the retrieval system and the host server for the purpose of presenting better results to the user. Anonymization, opt-out option and customization will be the central requirements, enforced through a protocol (see Section 6), to prevent any abuse that can result from sharing the information.

\section{Query-Aware Content Presentation}

Current state of the art Web servers do not take the query into consideration while presenting the content to the user. Lot of work has been reported on improving the architecture of Web servers for various applications \cite{1,2,4}. Many models are available to compare the architectures of the servers \cite{3}. \cite{5} discusses improving the performance of websites by using edge servers in Fog Computing Architecture. To the best of our knowledge there is no attempt to use the query by the host server to present the content to the user.

If a host server can present the content to the user based on the query, then it will be beneficial to both the user and the host organization. Suppose that user $A$ gives query ``popular movies in action genre + old'' and that $B$ gives ``popular movies in action genre + latest''. Let us assume that both the users get $Link1$ as their first link. We propose that the host server of $Link1$ should present different contents to each of them based on their query. In this case, the host server may present a list of old action movies (could be from other pages of the host server) to $A$ and a list of new action movies to $B$, in both case in addition to the content at the $Link1$.

For getting maximum benefit from this kind of architecture, the current Content Management Systems (CMS) like Drupal, Joomla, Django etc. may have to be redesigned to take user queries into account for presenting the final web page to be shown to the user. This will allow the host server (and the CMS) to play an active role in the process of content retrieval. Since the content provider knows much more about the content than the retrieval system, all that knowledge could be used to present dynamic query-aware content to the user.

\section{Feedback-Aware Retrieval}

A classical or bare-boned retrieval system \cite{7} only takes into account the query for retrieval. Some modern retrieval system go further and use the information available about the user for personalizing the results. However, they do not take into account the user's activity once the user has selected and visited one of the web pages. In the proposed architecture, anonymized information about the user's activity will be made available to the retrieval system. It will, thus, be possible to design algorithms that take this activity into account. Some work in this direction was proposed by \cite{6}. 

The details about this activity might include information such as the other links on the website that the user clicked on and the total time that the user spent on the website and on various pages. A retrieval system made aware of the feedback from the host server should, intuitively, perform better.

Some modern retrieval systems also provide additional links as part of the summary `snippet' while presenting the results of retrieval to the user. Such snippets can be better prepared with the suggested feedback from the host server.

Additionally, and importantly, the host server can provide extra information about its content as part of the feedback. This extra information will be based on the query and the knowledge of the content that is available to the host server. This will allow the content provider to have a say in the presentation of the snippet for the concerned website. The retrieval system may or may not use this information, depending on the retrieval and snippet preparation algorithm.

\section{Privacy and Customization}

Our proposal requires the retrieval system to share some information about the user and the query with the host server. It also requires the host server to provide feedback to the retrieval system based on the user's stay in the website after the user selected the website from the retrieval results. This extra sharing of information immediately raises the questions of privacy. If our proposal is implemented, its detailed version will need to include stringent requirements to address all the possible privacy concerns. We list below some of these requirements:

\begin{itemize}
\item The first such requirement is that the user's identity, even if known to the retrieval system, will not be revealed to the host server. Whatever information is shared with the host server will have to be strictly anonymized so as not to reveal the user's identity.
\item The second requirement is that only the relevant information will be shared. If we view this information as a list of attribute-value pairs, then only that subset of attribute-value pairs will be shared with the host server that the host server needs to know in order to better present its content.
\item The third requirement is that an opt-out option will be available to both the user and the host server. The user will be made aware of the sharing of information and the user will decide whether this sharing is to be allowed or not. The information will be shared only if the user explicitly agrees to it. In the default case, there will be no sharing. Similarly, the host server will decide whether to provide feedback to the retrieval system or not and the default will be the latter.
\item The fourth requirement is that both the user and the host server must be able to customize sharing of information. If they decide to share information, they will further be given the option to select the specific attributes that they are willing to share. For example, if the retrieval system knows about the user's location, age, gender and language, then the user may decide to share only location and language.
\item The user will have to be informed that the activity on the visited website may be used for providing feedback to the retrieval system. And the user will then decide whether and what part of the activity on the website can be used to provide feedback to the retrieval system.
\end{itemize}

As this proposal is worked out in more detail in future work, more such requirements might be identified and will also have to be addressed.

Even after addressing these issues, one concern still remains regarding the proposed architecture. Even if the shared information is anonymized and the host server does not know the identity of the user, the retrieval system may still know the identity and be able to connect the activity of the user on the visited website with the user's identity. This raises the question whether the retrieval system will come to acquire more knowledge about the user than is warranted. This may be a problematic ethical issue and requires further investigation.

\section{Retrieval Response Protocol}

There are many different kinds of retrieval systems. Similarly, there are many different kinds of host servers and content management systems. If there is to be a flow of information between them as suggested in the preceding sections, then it will have to be precisely regulated so that it is possible to implement systems without any conflict. This will require a well-defined and well-designed protocol. We call this the Retrieval Response Protocol (RRP).

The Retrieval Response Protocol will regulate the flow of information between the retrieval system and the host server. The protocol will be used to initiate, maintain and close a {\it retrieval session}. As soon as the user select one result from the results provided by the retrieval system in response to the user query, a retrieval session will be initiated. The ending of the session will perhaps have to be timeout based as there is no other way to know when the user has left the website.

During the time the session is alive, the retrieval system will first share the information about the user and the query with the host server. After that, based on the user's activity, the host server will provide the feedback to the retrieval system. All the activity during this session will be subject to the privacy and customization requirements and the protocol design will have to take this into account.

The protocol will have to be designed to regulate this retrieval session. We leave the design of this protocol for future work.

\section{Conclusion}

In the current information retrieval paradigm, the host does not use the query information for content presentation. The retrieval system does not know what happens after the user selects a retrieval result. And the host also does not have access to the information which is available to the retrieval system. We presented the outline of an architecture that addresses these issues. The aim is to provide a better search experience to the user through better presentation of the content based on the query and better retrieval results based on the feedback to the retrieval system from the host server. The retrieval system will share some information with the host server and the host server in turn will provide relevant feedback to the retrieval system based on the user's stay in the website. The host uses all the query related information for dynamic content presentation. This revised paradigm for information retrieval also introduces the issues of privacy which will have to be addressed stringently. It also needs a new protocol for content retrieval response, which we briefly described. This protocol will regulate the flow of information between the retrieval system and the host server subject to the privacy and customization requirements.

\bibliographystyle{plain}

\begin{thebibliography}{1}

\bibitem{1}
Ali Begen, Tankut Akgul, and Mark Baugher.
\newblock Watching video over the web: Part 1: Streaming protocols.
\newblock {\em IEEE Internet Computing}, 15(2):54--63, March 2011.

\bibitem{7}
Sergey Brin and Lawrence Page.
\newblock The anatomy of a large-scale hypertextual web search engine.
\newblock In {\em Proceedings of the Seventh International Conference on World
  Wide Web 7}, WWW7, pages 107--117, Amsterdam, The Netherlands, The
  Netherlands, 1998. Elsevier Science Publishers B. V.

\bibitem{3}
Ashif~S. Harji, Peter~A. Buhr, and Tim Brecht.
\newblock Comparing high-performance multi-core web-server architectures.
\newblock In {\em Proceedings of the 5th Annual International Systems and
  Storage Conference}, SYSTOR '12, pages 1:1--1:12, New York, NY, USA, 2012.
  ACM.

\bibitem{4}
Raoufehsadat Hashemian, Diwakar Krishnamurthy, Martin Arlitt, and Niklas
  Carlsson.
\newblock Improving the scalability of a multi-core web server.
\newblock In {\em Proceedings of the 4th ACM/SPEC International Conference on
  Performance Engineering}, ICPE '13, pages 161--172, New York, NY, USA, 2013.
  ACM.

\bibitem{6}
Xuehua Shen, Bin Tan, and ChengXiang Zhai.
\newblock Context-sensitive information retrieval using implicit feedback.
\newblock In {\em Proceedings of the 28th Annual International ACM SIGIR
  Conference on Research and Development in Information Retrieval}, SIGIR '05,
  pages 43--50, New York, NY, USA, 2005. ACM.

\bibitem{2}
Bryan Veal and Annie Foong.
\newblock Performance scalability of a multi-core web server.
\newblock In {\em Proceedings of the 3rd ACM/IEEE Symposium on Architecture for
  Networking and Communications Systems}, ANCS '07, pages 57--66, New York, NY,
  USA, 2007. ACM.

\bibitem{5}
Jiang Zhu, D.S. Chan, M.S. Prabhu, P.~Natarajan, Hao Hu, and F.~Bonomi.
\newblock Improving web sites performance using edge servers in fog computing
  architecture.
\newblock In {\em IEEE 7th International Symposium on Service Oriented System
  Engineering (SOSE)}, pages 320--323, March 2013.

\end{thebibliography}

\end{document}